\documentstyle[12pt,aps,prd]{revtex}

\def\beq{\begin{equation}}
\def\eeq{\end{equation}}
\def\ber{\begin{eqnarray}}
\def\eer{\end{eqnarray}}

\def\const{{\rm const}}
\def\ch{{\rm ch}}

\def\lsim{\ 
  \lower-1.2pt\vbox{\hbox{\rlap{$<$}\lower5pt\vbox{\hbox{$\sim$}}}}\ }
\def\gsim{\ 
  \lower-1.2pt\vbox{\hbox{\rlap{$>$}\lower5pt\vbox{\hbox{$\sim$}}}}\ }

\begin{document}
\sloppy

\title{Collapse of topological texture}

\author{O.~V.~Barabash$^{(a)}$ and Yu.~V.~Shtanov$^{(b)}$}  
\address{
$^{(a)}$Department of Physics, Shevchenko National University  
Kiev 252022, Ukraine \\
$^{(b)}$Bogolyubov Institute for Theoretical Physics 
Kiev 252143, Ukraine}

\date{6 July 1998}

\maketitle

\bigskip 

\begin{small}
We study analytically the process of a topological texture collapse 
in the approximation of a scaling ansatz in the nonlinear sigma-model.
In this approximation we show that in flat space-time topological texture 
eventually collapses while in the case of spatially flat expanding 
universe its fate depends on the rate of expansion. If the universe is 
inflationary, then there is a possibility that texture will expand 
eternally; in the case of exponential inflation the texture may also 
shrink or expand eternally to a finite limiting size, although this 
behavior is degenerate. In the case of power law noninflationary 
expansion topological texture eventually collapses. In a cold matter 
dominated universe we find that texture which is formed comoving with 
the universe expansion starts collapsing when its spatial size becomes 
comparable to the Hubble size, which result is in agreement with the 
previous considerations. In the nonlinear sigma-model approximation we 
consider also the final stage of the collapsing ellipsoidal topological 
texture. We show that during collapse of such a texture at least two of its 
principal dimensions shrink to zero in a similar way, so that their ratio 
remains finite. The third dimension may remain finite (collapse of 
{\em cigar\/} type), or it may also shrink to zero similar to the other two 
dimensions (collapse of {\em scaling\/} type), or shrink to zero similar to 
the product of the remaining two dimensions (collapse of {\em pancake\/} 
type). 

\bigskip

\noindent {PACS number(s): 11.27.+d, 98.80.Cq \hfill Preprint ITP-98-2E \par} 
{\hfill  hep-ph/9807291}
\end{small}

\newpage

\section{Introduction}

 Global textures are scalar field configurations that arise in 
theories in which global symmetry $G$ is spontaneously broken 
to $H$ in such a way that the vacuum manifold ${\cal M} = G/H$ has 
nontrivial third homotopy group $\pi_3({\cal M})$. 
Unlike in the case of lower dimensional {\em defects}---walls, strings, and 
monopoles---in a texture the scalar field can remain close to the vacuum 
manifold everywhere: textures do not possess cores. At the same time, 
texture configurations can be topologically nontrivial (in this case they 
are called {\em topological\/} textures). Such topological 
nontriviality guarantees that in the process of evolution
texture eventually collapses \cite{Derrick,Kibble}: there arises a 
shrinking spatial region in which its energy is concentrated. The details 
of such a phenomenon are discussed in the paper below. 

 Texture collapse events can set up the initial conditions for the structure 
formation in the early universe \cite{Turok}. This is one of the reasons why 
thorough study of such a process is of importance. The evolution of texture 
configurations has been studied numerically in \cite{STPR,PSB,BCL} and 
analytically in \cite{Perivolaropoulos,Sornborger}. The main issue that was 
under investigation in these papers is the fate of texture depending on the 
so-called winding number $w$, which is defined as the fraction of the vacuum 
manifold spanned by the texture configuration. Typical result was that under
each of the specific circumstances (flat space-time or expanding universe, 
spherical or nonspherical texture configuration, etc.) there exists a 
critical value $w_c < 1$ such that textures with $w > w_c$ collapse, while 
those with $w < w_c$ dissipate. In more recent paper \cite{SCP} the 
authors investigated numerically the influence of the geometry and topology
of the vacuum manifold on the dynamics of texture collapse. The dynamics 
of gravitational field during texture collapse were considered, for example,
in \cite{BV,DHJS}.

 In this paper we shall study analytically the case of a topological 
texture, for which $w$ is an integer, in a fixed background of flat 
space-time as well as in the expanding 
universe. Our goal will be to determine the conditions under which such a 
texture eventually collapses, and to describe the possible regimes of 
collapse. Thus our investigation may be regarded as complementary to the 
numerical simulations of \cite{STPR,PSB,BCL,SCP}.
Throughout this paper we shall use a simple approximation. 
We shall remain in frames of the nonlinear sigma-model [described by the 
Lagrangian (\ref{lag*}) below], and adopt a scaling ansatz that will reduce 
the complicated field evolution to the evolution of finite number of 
parameters. We hope that this approximation reflects the basic features of 
interest which we would like to reveal. First of all, in the preliminary 
Sec.~\ref{gen} we describe generation of cosmic texture during the symmetry 
breaking. In Sec.~\ref{flat} we describe topological
texture and show that in flat space-time its collapse is inevitable. In 
Sec.~\ref{frw} we study texture collapse in a spatially flat expanding 
universe and investigate the conditions under which collapse takes place. 
In Sec.~\ref{elli} we study the case of an ellipsoidal texture configuration
and investigate possible spatial shape of the collapse region in the 
vicinity of the moment of collapse. In Sec.~\ref{sum} we summarize our 
results. 

\section{Generation of cosmic texture} \label{gen}

 Consider a theory of the scalar fields
$\phi = \{\phi^A,\ A = 1,\ldots,N\}$ with the Lagrangian of type
\beq
L = \frac12 \sum_{A=1}^N \left(\partial \phi^A\right)^2 - V(\phi) \, ,
\label{lag}
\eeq
in which the potential $V(\phi)$ depends only on the value of
$\phi^2 = \sum_A \left(\phi^A\right)^2$ and has a global minimum at
$\phi^2 = \eta^2$. As an example one may consider 
\beq
V(\phi) = \frac g 4 \left(\phi^2 - \eta^2\right)^2 \, . \label{pot}
\eeq
The theory is invariant under the group 
$O(N)$ of transformations of the fields $\phi$.
The manifold of vacuum values of $\phi$ in this case has topology of  
$(N\!-\!1)$-dimensional sphere. 
As is well known \cite{Kibble} (see also \cite{Linde,VS,Peebles}), 
during thermal 
phase transition in which the symmetry becomes broken to $O(N\!-\!1)$ in the 
theory (\ref{lag}) there arise topological objects: if $N = 1$ these are 
domain walls, if $N = 2$---(global) strings, if $N = 3$---(global) monopoles, 
and, finally, if $N = 4$---the so-called (global) textures. These latter 
objects, textures, will be the subject of our investigation.
The above example Lagrangian (\ref{lag}) describes the situation 
of symmetry $O(4)$ spontaneously breaking to $O(3)$. However, the results 
of this and the following two sections will apply to a more general case of 
symmetry $G$ breaking to $H$ provided the third homotopy group of the 
vacuum manifold ${\cal M} = G/H$ is nontrivial. In Sec.~\ref{elli} we will 
use a spherically symmetric topological texture configuration, and the 
condition of existence of such configurations further restricts possible 
vacuum manifolds. For example, the theory with ${\cal M} = S^3$ will allow 
for spherically symmetric configurations, while the theory with 
${\cal M} = S^2$ will not (see \cite{SCP} for details). 

 Thermal phase transitions take place in the theory of hot expanding
universe. As a result of phase transition the scalar fields $\phi$
acquire their values close to the vacuum manifold, in our example it is a
three-sphere. Because during phase transition spatial regions that are
sufficiently remote from each other do not have enough time to exchange
signals the values of the scalar fields in such regions will be
uncorrelated. Thus as a result of phase transition there arises a
field configuration with the values of $\phi (x)$ lying close
to the vacuum manifold [$\phi^2 (x)$ close to $\eta^2$] 
but depending on the spatial point $x$ and changing on the
characteristic spatial scale $\xi \simeq t_{\rm ph}$, where 
$t_{\rm ph}$ is the characteristic time of the phase transition (the units
are chosen in which the speed of light is equal to unity). At every spatial
point $x$ the scalar fields tend to oscillate around the
closest vacuum value due to curvature of the potential $V(\phi)$ in the
``radial'' direction in the space of $\phi$. These oscillations fade away
rather rapidly as a result of the universe expansion as well as due to
excitation of the quanta of the fields $\phi$ and/or of the other fields
that interact with the fields $\phi$. In what follows we shall not be
interested in such oscillations, but turn to the space-time evolution 
of the vacuum values, which we denote as $\phi_* (x, t)$, closest
(in the vacuum manifold) to the values of $\phi (x, t)$.

 Since the fields $\phi_* (x, t)$ acquire values in the vacuum manifold, 
a three-sphere in our example,  
it is convenient to introduce arbitrary coordinates $\varphi = \{\varphi^a,\ 
a = 1,\ldots,3\}$ on this manifold and to describe the
evolution of interest in terms of the fields $\varphi (x, t)$. From
Eq.~(\ref{lag}) it then follows that the dynamics of the fields
$\varphi (x, t)$ is described by the Lagrangian of type 
\beq
L_* = \frac12 \sum_{a=1}^3 G_{ab} (\varphi)\, (\partial_\mu\varphi^a)
 (\partial_\nu\varphi^b) g^{\mu\nu} \, , \label{lag*}
\eeq
where
$G_{ab} (\varphi)$ is the metric components of the vacuum manifold in the
coordinates $\varphi$, and $g^{\mu\nu}$ is the inverse of the space-time 
metric tensor. The Lagrangian (\ref{lag*}) is that of a nonlinear 
sigma-model. In what follows we shall 
neglect the back-reaction of the scalar fields on the space-time metric
and regard the latter as fixed.

\section{Texture collapse in flat space-time} \label{flat}

 To begin with, we consider the evolution of cosmic texture, i.e.\@ of
a configuration of the fields $\varphi$, in flat space-time. Let the
field configuration be of finite total energy and, moreover, let there
exist a finite limit of $\varphi$ along any ray that goes to infinity
in the ordinary space. From finiteness of the energy it then follows 
\beq
\lim_{x \rightarrow \infty} \varphi (x) = \varphi_0 = \const \, .
\label{limit}
\eeq
We add infinitely remote point $x_0$ to the ordinary space and 
endow the space thus extended with the topology of a
three-sphere. In this topology the map $\varphi (x)$, defined also at the
point $x_0$ as $\varphi (x_0) = \varphi_0$, will be continuous. Being a
continuous map of a three-sphere (the extended ordinary space) into a
three-sphere (the vacuum manifold) it can belong to a nontrivial homotopy
class. If this happens the map $\varphi (x)$ cannot be continuously 
deformed to become a constant map. This feature will have interesting 
consequences. 

 The energy of a field configuration is given by the expression
\beq
E = \frac12 \int \sum_{ab} G_{ab} (\varphi) \left(\dot \varphi^a \dot
\varphi^b + \sum_i \nabla_i \varphi^a \nabla_i \varphi^b \right)d^3 x \, ,
\label{energy}
\eeq
in which the first part that depends on the time derivatives can be called
kinetic energy, and the second part---gradient energy. In the process of 
evolution the total energy is conserved. Given a field 
configuration $\varphi (x)$ in space the total energy will be minimized if
the time derivative $\dot \varphi$ is zero. Consider any of the homotopy
classes of configurations $\varphi (x)$. It is easy to see that
configurations from this class can acquire arbitrarily small energy
\cite{Derrick}. Indeed,
let $\varphi (x)$ be an arbitrary configuration from the homotopy class
considered. Putting $\dot \varphi \equiv 0$ we minimize the energy for the
configuration given. Then changing the configuration {\em continuously\/}
by the scaling transformation 
\beq
\varphi_\lambda (x) = \varphi (x / \lambda) \, , \ \ 
\lambda > 0 \, , \label{conf}
\eeq
which depends on the parameter $\lambda$, 
we can see from (\ref{energy}), that its {\em gradient\/} energy $\Pi$
changes as
\beq
\Pi_\lambda = \lambda\, \Pi \, , \label{poten}
\eeq
and acquires any positive value.

 During evolution of a field configuration with arbitrary initial
conditions $\{\varphi (x), \dot \varphi (x)\}$ its gradient energy
changes with time. From Eq.~(\ref{poten}) it is clear that its values 
can in principle approach zero arbitrarily closely. When $\Pi$ is close 
to zero then all over the space, with an exception of a bounded region,
the field values approach a constant $\varphi_0$. However, if the initial
configuration is topologically nontrivial then the map $\varphi (x, t)$
at any moment of time must cover the whole three-sphere vacuum manifold. 
This means that in the spatial region in which the fields $\phi_*$ change
substantially (we are speaking now in terms of original fields $\phi$) 
their characteristic change is at least of order $\eta$. If by $R$ we denote
the characteristic linear dimension of the spatial region of variation of
the fields $\phi_*$ then the value of the gradient energy 
\beq
\Pi \sim \eta^2 R \,
\eeq
decreases with the decrease of $R$. Thus energetically it is favourable
for the value $R$ to decrease. At the same time the characteristic
{\em density\/} $\rho_\Pi$ of the gradient energy in the region of field
variation behaves as
\beq
\rho_\Pi \sim \left({\eta \over R}\right)^2 \, ,
\eeq
and grows infinitely with the decrease of $R$. Such a situation, when the
value of $R$ decreases in time and the gradient energy density increases
in the region of field variation, is characterized as collapse of cosmic
texture. 

 Now becomes clear the role which in our reasoning is played by the
topological
nontriviality of the initial field configuration.\footnote{A topologically 
nontrivial field configuration is usually called {\em topological texture}.} 
Were this configuraion
topologically trivial, then evolving continuously it could in principle
approach a constant $\varphi_0$ {\em uniformly\/} in the space (the process 
which is sometimes called {\em dissipation}). 
Topological nontriviality
guarantees that the map $\varphi (x, t)$ at every moment of time covers
the {\em whole\/} three-sphere of the vacuum manifold. In such a case the
gradient energy of the field configuration can decrease to zero only as
a result of the collapse described above. If the texture is not topological, 
then it will collapse or it will dissipate without collapse depending to a 
large extent on the so-called winding number $w$ which is the fraction of 
the vacuum manifold covered by the texture. The corresponding analysis 
was performed in the papers \cite{STPR,PSB,BCL,Perivolaropoulos,Sornborger}. 

 In reality the energy density of a topological texture cannot increase
unbounded. When the size of
the collapse region becomes so small that the gradient energy density
$\rho_\Pi$ is of the same order as the value $V(0)$ of the potential
in its local maximum [we take this maximum to be at $\phi = 0$ as in the
example (\ref{pot})] the field dynamics becomes more complicated: the
field values depart from the vacuum manifold rather far, in particular,
they can acquire the values $\phi = 0$ and overcome the potential
barrier that separates different homotopy classes. The texture acquires
the possibility to ``unwind'' itself. After this nothing prevents it from
evolving uniformly to a constant $\phi_0$ (although before this happens 
several recollapses may occur, see \cite{PSB}). From the above 
discussion it is clear that ``unwinding'' occurs when 
\beq
\rho_\Pi \sim \left({\eta \over R}\right)^2 \sim V(0) \, ,
\eeq
which for the potential (\ref{pot}) gives the estimate 
\beq
R \sim {1 \over \eta \sqrt g} \, .
\eeq

 In order to describe qualitatively the dynamics of texture collapse in
this work we shall adopt the scaling approximation. In the case
of flat space-time we place the origin of the space-time coordinates at
the centre of mass of the initial field configuration and
choose the inertial coordinate system in which this centre is at rest.
After that we consider the family of configurations (\ref{conf})
parametrized by $\lambda$. We shall take that the texture evolution proceeds
within this family. Of course, this is not so in reality, but
we hope this can serve as a reasonable approximation to the actual
evolution. Note that an exact solution
of the scaling type was found in the work \cite{TS} for the collapse of
a spherically symmetric topological texture with infinite total energy. 

 For the family of configurations (\ref{conf}) the action as a functional
of $\lambda (t)$ has the form 
\beq
S[\lambda (t)] \equiv S[\varphi_{\lambda(t)} (x)] =
\int \left(A \lambda \dot \lambda^2 - B \lambda \right) dt \, ,
\eeq
where positive constants $A$ and $B$ are given by 
\ber
&{}& A = \frac12 \int G_{ab} (\varphi) (x \cdot \nabla
\varphi^a) (x \cdot \nabla \varphi^b) d^3 x \, , \label{A} \\
&{}& B = \frac12 \int G_{ab} (\varphi) \nabla \varphi^a \cdot \nabla
\varphi^b\, d^3 x \, . \label{B}
\eer
Note that $\varphi = \varphi (x)$ is the initial field configuration (when
$\lambda = 1$), and the dot denotes the scalar product in the
three-dimensional euclidean metric. From the energy conservation law 
\beq
E = A \lambda \dot \lambda^2 + B \lambda = \const
\eeq
we get the equation 
\beq
\dot \lambda = \pm \sqrt{E - B \lambda \over A \lambda} \, ,
\eeq
which can be integrated. If initially $\dot \lambda < 0$ then the texture
starts shrinking immediately and eventually collapses. The case when 
initially $\dot \lambda > 0$
also leads to eventual collapse. Indeed, the value of $\lambda$ cannot
increase unbounded, and at the turning point $\lambda = E / B$ we have
$\ddot \lambda = - B^2 / 2AE \ne 0$ so that texture cannot eternally expand.
Thus in the scaling approximation considered the texture eventually
collapses under arbitrary initial conditions. The time of collapse from the 
turning point $\lambda = E / B$ till $\lambda = 0$ is estimated as  
\beq
t_{\rm coll} = \int\limits_0^{E/B} d \lambda \sqrt{A \lambda \over 
E - B \lambda} = {\pi E \sqrt A \over 2 B^{3/2}} \sim R_*\, , 
\eeq
where $R_*$ is the spatial size of the texture at the turning point. 

 We end this section by the following remark. It is impossible that
as a result of a phase transition in the {\em infinite\/} space there
arises a field configuration of {\em finite\/} energy. Actually it is only
the average energy {\em density\/} that will be finite. Nevertheless,
in the process of evolution the field configuration tends to become
constant in as large spatial regions as possible. Thus spatial islands may
arise in which fields change substantially, surrounded by the sea of
practically constant fields. It is such island regions to which our
analysis will then be applicable.

\section{Texture collapse in a spatially flat expanding universe}
\label{frw}

 As it was shown in the previous section in the scaling
approximation, in flat space-time any topological texture eventually
collapses. In an expanding universe texture collapse is not inevitable.
In this section we investigate the conditions under which it takes place. 

 Consider evolution of a cosmic texture in a spatially flat expanding
universe. As we did in the previous section, we shall use the scaling
approximation, that is, we restrict texture configurations to the family
(\ref{conf}) with $\varphi (x)$ being the initial configuration. Moreover, 
in this section we shall
regard the motion of the texture centre of mass with respect to the
cosmological background to be negligible. Then the action as a functional
of $\lambda (t)$ will be 
\beq
S[\lambda (t)] = \int \left(A a^3 \lambda \dot \lambda^2 - B a \lambda
\right) dt \, ,
\eeq
where $a (t)$ is the universe scale factor, and the constants $A$ and $B$
are given respectively by (\ref{A}) and (\ref{B}).

 Physical size of the texture is proportional to the product
$a \lambda$. We shall proceed therefore to a new variable 
\beq
\mu = a \lambda \, .
\eeq
The Lagrangian in terms of this variable acquires the form
\beq
L = A \mu \left(\dot \mu - H \mu \right)^2 - B \mu \, ,
\eeq
and the equations of motion are 
\beq
2 \mu \ddot \mu + \dot \mu^2 - \left(2 \dot H + 3 H^2 \right) \mu^2
+ b = 0 \, , \label{eqmu}
\eeq
where $H \equiv \dot a / a$ is the Hubble parameter, and $b = B / A$. One
can get rid of the first time derivative in Eq.~(\ref{eqmu}) by making
another change of variable 
\beq
\nu = \mu^{3/2} \, . \label{change}
\eeq
For the value $\nu (t)$ we then get the equation 
\beq
\frac43 \ddot \nu = f (t) \nu - b \nu^{-1/3} \, , \label{eqnu}
\eeq
where
\beq
f (t) \equiv 2 \dot H + 3 H^2 \, .
\eeq

 Consider one important class of the universe expansion laws
\beq
a (t) \propto t^\alpha \, , \ \ \ \alpha > 0 \, , 
\eeq
which is realized in the standard cosmological model. Thus $\alpha = 1/2$
if the universe is radiation dominated, and $\alpha = 2/3$ if it is
dominated by nonrelativistic matter (``dust''). For this class we have 
\beq
f (t) = {\alpha (3 \alpha - 2) \over t^2} = {\beta \over t^2}\, , \label{f}
\eeq
in which we made notation $\beta = \alpha (3 \alpha -2)$. If $\alpha \le 1$
then the texture always eventually collapses, that is, $\nu (t)$ acquires
the value of zero in a finite time. Indeed, for $\alpha \le 2/3$ we have
$f (t) \le 0$, and from Eq.~(\ref{eqnu}) it follows immediately that the
texture
eventually collapses. Now let $2/3 < \alpha < 1$, and let us assume that
there exists solution $\nu (t)$ which never becomes zero in the future. 
This solution should be bounded from above by the one which is obtained by 
dropping the second term from the right-hand side of Eq.~(\ref{eqnu}) and 
imposing the same initial conditions. Thus at $t \rightarrow \infty$ we 
should have $\nu (t) t^{-3/2} \rightarrow 0$. But for such a behavior of 
$\nu (t)$ at sufficiently large values of $t$ the second term in the 
right-hand side of (\ref{eqnu}) should become 
dominating, so that the texture expansion must be followed by its
contraction and collapse, what contradicts the assumption made. Somewhat
more lengthy analysis (see Appendix~\ref{AA}) shows that texture collapse 
is inevitable also for $\alpha = 1$ (Milne's cosmology). 

 If $\alpha > 1$ then also $\beta > 1$; in this case there exists solution 
\beq
\nu (t) = \left({\beta \over \beta - 1}\right)^{3/4} t^{3/2} \, ,
\eeq
which corresponds to eternally expanding texture.

 Consider now two important cases in which Eq.~(\ref{eqmu}) can be easily
integrated.

\subsection{The case of $a (t) \propto t^{2/3}$}

 In the case of $a (t) \propto t^{2/3}$ we have $f = 2 \dot H + 3 H^2 \equiv 0$. 
The equation (\ref{eqmu}) for $\mu (t)$ in this case is easily integrated and 
the result can be put in a parametric form 
\ber
\mu &=& c \sqrt b (1 - \cos \tau) \, , \label{cyc1} \\
t &=& c (\tau_0 + \tau - \sin \tau) \, , \label{cyc2}
\eer
where $c$ and $\tau_0$ are the integration constants and $\tau$ the parameter. 
The solution describes a cycloid. The beginning of texture collapse, i.e.\@  
the turning point, corresponds to $\tau = \pi$. The physical values at this 
moment will be labelled by an asterisk. At this moment the ratio of 
the texture size $R_* = r \mu_*$ to the Hubble radius 
$H_*^{-1} = 3t_*/2$ is 
\beq
R_* H_* = r \mu_* H_* = {4 r \sqrt b \over 3 \left(\pi + \tau_0 \right)} \, ,
\label{rati}
\eeq
where $r$ is the coordinate size of the configuration $\phi (x)$.

 Let us estimate the value of $b$. Using (\ref{A}) and (\ref{B})
we obtain 
\beq
A \sim \eta^2 r^3\, , \ \ \ B \sim \eta^2 r \, ,
\eeq
and
\beq
b = {B \over A} \sim r^{-2} \, .
\label{bstar}
\eeq
Thus for the value (\ref{rati}) we have 
\beq
R_* H_* \sim {1 \over \pi + \tau_0} \, . 
\eeq

 Let texture form
at the moment of time $t_0$ with its characteristic size $R_0$ at this
moment much larger than the Hubble size, 
\beq
R_0 H_0 \gg  1 \, .  \label{inits}
\eeq
At the moment of its generation let texture expand with the universe,
that is,  
$\dot \lambda (t_0) = 0$, and, therefore, 
\beq
\dot \mu (t_0) = H_0 \mu_0 \, ,  \label{initmu}
\eeq
where the subscript ``{\small 0}'' labels various quantities at the moment 
$t_0$. Then it is easy to see that the integration constant $\tau_0$ in 
(\ref{cyc2}) must be much smaller than unity, hence 
\beq
R_* H_* \sim {1 \over \pi}\, .
\eeq
This result implies that the texture starts collapsing when its linear
dimension compares with the Hubble size, which is in agreement with the
reasonable expectations (see \cite{Turok}).

\subsection{The case $a (t) \propto \exp (Ht), \ H \equiv \const$}

 According to cosmological inflation scenarios \cite{Linde} in the very
early universe there might have been a period of (almost) exponential 
expansion during which
\beq
a (t) \propto \exp (Ht)\, , \ \ \ H \approx \const \, .
\eeq
Textures also might form during this period as a result of spontaneous
symmetry breaking. We shall investigate their possible subsequent evolution 
in the scaling approximation. 

 In the case of $H \equiv \const$ the equation (\ref{eqmu}) can be integrated
with the result 
\beq
\dot \mu = \pm \sqrt{H^2 \mu^2 - b + C/\mu} \, ,  \label{dotmu}
\eeq
where $C$ is the integration constant, and $b = B/A$ as before. Depending
on the value of $C$ the cubic three-nom
\beq
H^2 \mu^3 - b \mu + C  \label{nom} 
\eeq
can:
\newcounter{num}
\begin{list}
{(\roman{num})}{\usecounter{num}\setlength{\rightmargin}{\leftmargin}}
\item have no positive roots; this case is realized if $C > C_0$;
\item have only one positive root; this case is realized if $C = C_0$
or $C \le 0$;
\item have two positive roots: this case is realized if $0 < C < C_0$.
\end{list}
The constant $C_0$ above is equal to 
\beq
C_0 = {2 \over H} \left({b \over 3}\right)^{3/2} \, .
\eeq

 In the case (i) the texture expands unbounded or monotonically
collapses depending on the sign in (\ref{dotmu}). In the case (iii) 
the cubic three-nom (\ref{nom}) has two positive roots $\mu_1$ and $\mu_2
> \mu_1$. In this case the region $\mu_1 < \mu < \mu_2$ of the values of 
$\mu$ is forbidden. If initially $0 < \mu < \mu_1$, then for the initial time
derivative $\dot \mu < 0$ the texture collapses monotonically. If, however,
initially $\dot \mu > 0$, then the texture first expands to the value
of $\mu = \mu_1$, after which the derivative $\dot \mu$ changes its sign
and the texture collapses. The expansion to the point of $\mu = \mu_1$
proceeds in a finite time since in the vicinity of this point we have
$\dot \mu \propto \sqrt{\mu_1 - \mu}$. The case $\mu > \mu_2$ is analysed
in a similar way. In this case the texture eventually expands unbounded
independently of the sign of the initial time derivative $\dot \mu$.
The case (ii) for $C \le 0$ is totally analogous to the case (iii)  
in the region $\mu > \mu_2$. In a particular case $C = 0$ the solution
is expressed through elementary functions 
\beq
\mu (t) = {\sqrt b \over H} \ch \left[H (t - t_0) \right] \, .
\eeq
Consider the remaining case of $C = C_0$. The positive root of the three-nom
(\ref{nom}) in this case is twofold degenerate,
$\mu_1 = \mu_2 = {\sqrt{b/3} \over H}$. Let initially we have $\mu < \mu_1$.
Then, if $\dot \mu < 0$ the texture will collapse monotonically. If
$\dot \mu > 0$ it will expand to the size $\mu = \mu_1$. Since the root
$\mu_1$ is twofold degenerate, in its vicinity we will have $\dot \mu
\propto (\mu_1 - \mu)$ and the value $\mu = \mu_1$ is attained in infinite
time. Thus for $\mu < \mu_1$ and $\dot \mu > 0$ the texture expands
eternally approaching the value of $\mu = \mu_1$. In the case 
$\mu > \mu_1$ the texture will expand unbounded if $\dot \mu > 0$, and
contract eternally approaching the value of $\mu = \mu_1$ if $\dot \mu < 0$. 
Note that such a behavior is also highly unlikely since in order for it to 
be realized the values of $\mu$ and of its time derivative $\dot \mu$ are to 
be fine tuned from the beginning.

\section{Collapse of an ellipsoidal texture} \label{elli}

 In this section we consider collapse of an ellipsoidal texture configuration
[see Eq.~(\ref{ell}) below] in a theory which allows for spherically symmetric
topological texture configurations (see \cite{SCP} for details). For instance, 
this can be the theory with Lagrangian (\ref{lag}) in which the symmetry $O(4)$
is spontaneously broken to $O(3)$. In the vicinity of the collapse point the 
cosmological curvature of the space-time
geometry can be disregarded, and we shall take the space-time to be flat.
As to the scaling ansatz we set 
\beq
\varphi_{\{\lambda_i\}} (x_1, x_2, x_3) = \chi \left({x_1 \over
\lambda_1}, {x_2 \over \lambda_2}, {x_3 \over \lambda_3} \right) 
\label{ell} \, ,
\eeq
where $x_i$ are the spatial Cartesian coordinates, $\lambda_i$ are parameters
analogous to the $\lambda$ in Eq.~(\ref{conf}), and $\chi (x)$ is a spherically 
symmetric topological configuration, such that $\lim\limits_{x \rightarrow \infty}
\chi (x) = \varphi_0$. The parameters $\lambda_i$ determine the dimensions of 
the ellipsoidal 
configuration (\ref{ell}). By adopting the approximation according to which
the texture evolution proceeds within the three-parameter
family (\ref{ell}) we obtain after straightforward calculation the action
as a functional of $\{\lambda_i (t)\}$:
\beq
S[\{\lambda_i (t)\}] = \int \left(\sum_{ij}A_{ij} h_i h_j - B \sum_i
\lambda_i^{-2} \right) \prod_k \lambda_k\, dt \, , \label{action}
\eeq
where 
\beq
h_i = {\dot \lambda_i \over \lambda_i} \, ,
\eeq
and, in view of the spherical symmetry imposed, 
\beq
A_{ij} = (3A_1 + 2A_2)\, \delta_{ij} + (A_2 - A_1) \, ,
\eeq \beq
A_1 = \frac{1}{60} \int G_{ab}(\chi) \sum_{ij} \nabla_i \chi^a \nabla_j \chi^b
\left(r^2 \delta_{ij} - x_i x_j \right) d^3 x > 0 \, , \label{A1}
\eeq \beq 
A_2 = \frac{1}{30} \int G_{ab}(\chi) \sum_{ij} \nabla_i \chi^a \nabla_j \chi^b
x_i x_j\, d^3 x > 0 \, , \label{A2}
\eeq \beq 
B = \frac16 \int G_{ab} (\chi) \sum_i 
\nabla_i \chi^a \nabla_i \chi^b \, d^3 x > 0 \, . \label{Bnew}
\eeq
Above we made a notation $r^2 = \sum_i x_i^2$. In (\ref{A1})--(\ref{Bnew}) 
$\chi = \chi (x)$---the spherically symmetric configuration. 
Note that the constant $B$ introduced in (\ref{Bnew}) should not be confused 
with the analogous constant in Sections~\ref{flat} and \ref{frw}. 

 From the action (\ref{action}) we obtain the equations of motion
\beq
10 \dot h_k + 10 h_k \sum_i h_i - s \sum_i h_i^2 - p \left(\sum_i h_i \right)^2
- 10 b \lambda_k^{-2} + b(4 + q) \sum_i \lambda_i^{-2} = 0 \, , \label{dothk}
\eeq
where 
\beq
s = 3q + 2\, , \ \ p = 1 - q \, , \ \ q = {A_1 \over A_2} \, , \ \ 
b = {B \over s A_2} \, , 
\eeq
and also the expression for the total energy
\beq
E = A_2 \prod_j \lambda_j \left[s \sum_i h_i^2 + p \left(\sum_i h_i\right)^2
+ s b \sum_i \lambda_i^{-2} \right] \, . \label{energy-el}
\eeq

 Let $t_0$ denote the time moment of collapse, when one of the $\lambda_i$ 
turns to zero, and let this be $\lambda_1$. From the expression 
(\ref{energy-el}) 
for the energy, and from the energy conservation law, it then follows that 
at least one of the remaining two parameters, say $\lambda_2$, must also 
turn to zero at this moment of time. If $\lambda_3$ remains finite then 
from the energy conservation law it also follows that
\beq
\lim_{t \rightarrow t_0} {\lambda_1 \over \lambda_2} = \zeta\ne 0 \, , 
\label{ratio}
\eeq
and that the values of 
$\dot \lambda_1$ and $\dot \lambda_2$ must be bounded in the vicinity of $t_0$. 
It can be shown (see Appendix~\ref{AB}) that in the case considered in fact 
$\zeta = 1$ in Eq.~(\ref{ratio}).  

 Now consider the case when at $t = t_0$ all three parameters $\lambda_i$ 
vanish. Let us look for the solutions which in the vicinity of $t = t_0$ have 
asymptotic behavior 
\beq
\lambda_i \sim \Lambda_i \left(t_0 - t\right)^{\alpha_i} \, , \ \ \ 
i = 1,2,3 \, , \label{an}
\eeq
with positive constants $\Lambda_i$ and $\alpha_i$. Then in the vicinity of 
$t = t_0$ 
\beq
h_i \sim - {\alpha_i \over t_0 - t}\, , \ \ \ \dot h_i \sim 
- {\alpha_i \over (t_0 - t)^2} \, . \label{hs}
\eeq
Taking the sum of Eq.~(\ref{dothk}) over the index $k$, multiplying the result 
by $(t_0 - t)^2$, then taking the limit of $t \rightarrow t_0$ and using 
(\ref{an}), (\ref{hs}) we obtain 
\beq
- 10 \sum_i \alpha_i + (10 - 3p) \left(\sum_i \alpha_i\right)^2 - 3 s 
\sum_i \alpha_i^2 + s b \lim_{t \rightarrow t_0} \left[(t_0 - t)^2 \sum_i 
\Lambda_i^{-2} (t_0 - t)^{-2\alpha_i} \right] = 0 \, . \label{d}
\eeq
In order that the energy 
(\ref{energy-el}) remains nonzero and finite in the limit of 
$t \rightarrow t_0$, and also that the limit in (\ref{d}) is finite, 
it is necessary that 
\beq
\sum_i \alpha_i = 2 \, ; \ \ \mbox{and} \ \ \alpha_i \le 1 \ \ \mbox{for all} 
\ \ i \, . 
\eeq
 
 If $\alpha_i < 1$ for all $i$ then from (\ref{d}) we find the unique solution 
\beq
\alpha_i = \frac23 \, \ \ \mbox{for all} \ \ i \, . 
\eeq
It is easy to check that the asymptotic expressions (\ref{an}) with these 
values of $\alpha_i$ satisfy the system (\ref{dothk}) to the leading 
approximation. We thus recover the scaling collapse which we studied in 
Sec.~\ref{flat}. 
In this case 
\beq
\lim_{t \rightarrow t_0} (\lambda_1 : \lambda_2 : \lambda_3) = 
\Lambda_1 : \Lambda_2 : \Lambda_3 \, . \label{lim3}
\eeq

 If, for example, $\alpha_3 = 1$ then using the system of equations 
(\ref{dothk}) we obtain the solution 
\beq
\alpha_1 = \alpha_2 = \frac12 \, , \ \ \ \Lambda_3 = \sqrt{2b} \, . 
\eeq
In this case 
\beq
\lim_{t \rightarrow t_0} {\lambda_1 \over \lambda_2} = {\Lambda_1 \over 
\Lambda_2} \, , \ \ \ \ \lim_{t \rightarrow t_0} {\lambda_1 \lambda_2 \over 
\lambda_3} = {\Lambda_1 \Lambda_2 \over \Lambda_3} \, . \label{lim2} 
\eeq

 Thus an ellipsoidal texture always collapses similar in two directions. 
In the third direction its dimension can behave in three possible ways: 
(i)~remain finite, then Eq.~(\ref{ratio}) is valid with $\zeta = 1$, 
(ii)~shrink similar to the other two dimensions, so that Eq.~(\ref{lim3}) 
is valid, (iii)~shrink faster, so that Eq.~(\ref{lim2}) is satisfied. 

 Note that our analysis was performed in frames of the nonlinear sigma-model 
with the Lagrangian (\ref{lag*}). As it was emphasized in Sec.~\ref{flat} 
this approximation breaks down as soon as one of the texture spatial 
dimensions becomes as small as 
\beq
R \sim {1 \over \eta \sqrt g } \, , 
\eeq
where, we remember, $g$ is the coupling constant in (\ref{pot}). 
At this moment texture unwinding is expected to occur. 

 It is worth noting that the solution in the form (\ref{an}) becomes 
exact if one neglects the last two terms in Eq.~(\ref{dothk}). One may raise 
the issue about the role of these two terms. It is easy to see that in all 
the cases considered above their influence will be in the direction of 
reducing the texture spherical asymmetry. This is indeed observed 
in numerical simulations \cite{STPR,PSB}. However, simulations in 
\cite{STPR,PSB} are 
not limited to the case of nonlinear sigma-model (\ref{lag*}), they are based 
on the original Lagrangian (\ref{lag}), which means that field variation in 
radial direction (in the space of fields) may also be of significance in 
producing the effect discussed in \cite{STPR,PSB}.

\section{Summary} \label{sum}

 In this paper we studied analytically the process of topological texture 
collapse using the approximation of scaling ansatz in the nonlinear 
sigma-model. Our investigation may be regarded as complementary to the 
numerical simulations of 
\cite{STPR,PSB,BCL,SCP}. We have seen that in flat space-time topological
texture eventually collapses while in the case of spatially flat expanding 
universe its fate depends on the rate of expansion. If the universe is 
inflationary, $a(t) \propto t^\alpha$ with $\alpha > 1$, or $a(t) \propto
\exp (Ht)$, then there is a possibility that texture will expand eternally.
In the case of exponential inflation there is also a possibility of eternal 
shrinking or expansion to a finite limiting size, however, this solution 
is degenerate. In the case of power law universe expansion, $a(t) \propto 
t^\alpha$, with $0 < \alpha \le 1$ topological 
texture eventually collapses. In the case of cold matter dominated universe, 
$a(t) \propto t^{2/3}$, we have seen that texture which is formed comoving
with the universe expansion 
starts collapsing when its spatial size becomes comparable to the Hubble 
size, which is in agreement with the previous considerations. We considered 
also the final stage of collapsing ellipsoidal topological texture. Remaining 
in the approximation of the nonlinear sigma-model with the Lagrangian 
(\ref{lag*}) we 
have seen that during collapse of such a texture at least two of its principal
dimensions shrink to zero in a similar way, so that their ratio 
remains finite. The third dimension may remain finite [collapse of {\em cigar\/}
type], or it may also shrink to zero similar to the other two dimensions 
[collapse of {\em scaling\/} type, Eq.~(\ref{lim3})], or shrink to 
zero similar to the product of the remaining two dimensions 
[collapse of {\em pancake\/} type, Eq.~(\ref{lim2})]. 

\bigskip

\section*{Acknowledgments}

We are grateful to R.~H.~Brandenberger for reading the manuscript and 
making several useful suggestions. 
The work of Yu.~S. was supported in part by the Foundation of Fundamental 
Research of the Ministry of Science and Technology of Ukraine under the 
grant No~2.5.1/003.

\bigskip 

\appendix

\section{The case of Milne space-time} \label{AA}

 In this Appendix we shall investigate Eq.~(\ref{eqnu}), in the case of 
Milne's cosmology: 
\beq
a (t) \propto t \, , 
\eeq
for which the function $f (t)$
is given by the expression (\ref{f}) with $\beta = 1$. We shall show that
in this case any solution describes eventual collapse, that is, $\nu (t)$
attains the value of zero. In Eq.~(\ref{eqnu}) with $\beta = 1$ we 
make substitution 
\beq
t = \sqrt{\tau \over b}\, , \ \ \ \nu = \sigma \tau^{-1/4} \, . \label{new}
\eeq
As a result we get the equation for $\sigma (\tau)$:
\beq
\ddot \sigma = - \frac{3}{16} \left(\sigma \tau^2 \right)^{-1/3} \, ,
\label{eqv}
\eeq
in which the dot denotes the derivative with respect to $\tau$.

 Assume that there exists solution $\nu (\tau)$ without collapse. Then from
the equations (\ref{eqnu}) and (\ref{new}) it follows that 
\beq
\lim_{\tau \rightarrow \infty} \sigma (\tau) = \infty \, . \label{limv}
\eeq
According to Eq.~(\ref{eqv}) the derivative 
$\dot \sigma (\tau)$ monotonically decreases, and according to 
Eq.~(\ref{limv}) it is bounded by zero from below. Hence, there exists
a finite limit
\beq
\lim_{\tau \rightarrow \infty} \dot \sigma (\tau) = v \ge 0 \, . 
\label{limdv}
\eeq
If we assume that $v > 0$, then asymptotically $\sigma \sim v \tau$, and 
integration of Eq.~(\ref{eqv}) would yield $\dot \sigma \sim \const \cdot 
\log \tau$ at
large $\tau$ which is incompatible with Eq.~(\ref{limdv}). Therefore in the 
equality (\ref{limdv}) there must be $v = 0$.

 Using the de~l'Hospital rule we calculate the limit 
\beq
\lim_{\tau \rightarrow \infty} {\sigma (\tau) \over \tau} = 
\lim_{\tau \rightarrow \infty} \dot \sigma (\tau) = 0 \, . \label{limvot}
\eeq
Thus 
\beq
\sigma (\tau) = \tau y (\tau) \, , \ \ \ \lim_{\tau \rightarrow \infty} 
y (\tau) = 0 \, , \label{y}
\eeq
whence, with Eq.~(\ref{limvot}) taken into account, it follows 
\beq
\lim_{\tau \rightarrow \infty} \tau \dot y (\tau) = 0 \, . \label{limtdy}
\eeq

 From Eqs.~(\ref{eqv}), (\ref{y}) we obtain the equation for $y (\tau)$:
\beq
\tau^2 \ddot y + 2 \tau \dot y = - {3 \over 16 y^{1/3}} \, ,
\eeq
from which with (\ref{y}), (\ref{limtdy}) taken into account it follows
\beq
\lim_{\tau \rightarrow \infty} \tau^2 \ddot y = - \infty \, . 
\label{limtddy}
\eeq
Applying once again the de~l'Hospital rule and using (\ref{limtdy}),
(\ref{limtddy}), we get 
\beq
0 = \lim_{\tau \rightarrow \infty} \tau \dot y (\tau) = 
\lim_{\tau \rightarrow \infty} {\dot y (\tau) \over (1/\tau)} = 
\lim_{\tau \rightarrow \infty} {\ddot y (\tau) \over (1/\tau) \dot{}} = 
- \lim_{\tau \rightarrow \infty} \tau^2 \ddot y (\tau) = 
\infty \, . \label{limfin}
\eeq

 Therefore, the initial assumption of the absence of collapse has lead to 
contradiction. Thereby it is proven, within the scaling approximation, that in 
the universe that expands as $a (t) \propto t$ any texture eventually collapses.

\section{The value of $\zeta$} \label{AB}

 Here we show that the value of $\zeta$ in Eq.~(\ref{ratio}) is equal to one. 
It is reasonable to assume that the functions $\lambda_i (t)$, $i = 1, 2, 3$, 
together with their first and second derivatives behave monotonically in the 
vicinity of the collapse moment of time $t_0$. This guarantees the existence
of all the limits below. As it was mentioned after Eq.~(\ref{ratio}), the 
values of $\dot \lambda_i$, $i = 1, 2$, are bounded 
in the vicinity of the collapse moment $t_0$, hence 
\beq
\lim_{t \rightarrow t_0} \dot \lambda_i = \omega_i \, , \ \ i = 1, 2 \, . 
\label{limdl}
\eeq
By the de~l'Hospital rule we have 
\beq
\lim_{t \rightarrow t_0} {\lambda_1 \over \lambda_2} = \lim_{t \rightarrow t_0} 
{\dot \lambda_1 \over \dot \lambda_2} = \zeta \, . \label{limratio}
\eeq
Now applying the de~l'Hospital rule in the third equality just below we get  
\beq
\omega_i \omega_j = \lim_{t \rightarrow t_0} \left(\dot \lambda_i \dot 
\lambda_j \right) = \lim_{t \rightarrow t_0} \left({\lambda_i \dot \lambda_j 
\over t - t_0}\right) = \lim_{t \rightarrow t_0} \left(\dot \lambda_i \dot 
\lambda_j + \lambda_i \ddot \lambda_j \right) = \omega_i \omega_j + 
\lim_{t \rightarrow t_0} \left(\lambda_i \ddot \lambda_j \right) \, ,
\eeq
hence  
\beq
\lim_{t \rightarrow t_0} \left(\lambda_i \ddot \lambda_j \right) = 0 \, .
\label{long} 
\eeq
The existence of the last limit is guaranteed by the assumed monotonic 
behavior of the functions' second derivatives. Using (\ref{limdl}) and 
(\ref{long}) we obtain 
\beq
\lim_{t \rightarrow t_0} \lambda_i h_j = \omega_i \, , \ \ \  
\lim_{t \rightarrow t_0} \lambda_i^2 \dot h_j = - \omega_i^2 \, , \label{fin} 
\eeq
where, we remember, $h_k \equiv \dot \lambda_k / \lambda_k$, and the indices
$i, j$ independently of each other take on the values $1, 2$. From 
the law of conservation of energy (\ref{energy-el}) it also follows that the 
values of $\lambda_i h_3$, $i = 1, 2$, have finite limits as $t \rightarrow 
t_0$. 

 Finally, we take the equations of motion (\ref{dothk}) with $k = 1$ and 
$k = 2$, multiply each of them by $\lambda_1^2$ and take the limit of 
$t \rightarrow t_0$. Using (\ref{limratio}), (\ref{fin}) and finiteness of 
the product $\lambda_1 h_3$ we obtain $\zeta = 1$.

\end{document}